\documentclass[twocolumn,letter,appendixfloats]{aastex61}

\usepackage[fleqn]{amsmath}
\makeatletter
\newcommand*{\rom}[1]{\expandafter\@slowromancap\romannumeral #1@}
\makeatother

\shorttitle{NIR P-L relations of Miras in the LMC}
\shortauthors{Yuan et al.}

\begin{document}
\title{Large Magellanic Cloud Near-Infrared Synoptic Survey. V.\\Period-Luminosity Relations of Miras.}

\author[0000-0001-9420-6525]{Wenlong Yuan}
\affiliation{George P.~and Cynthia W.~Mitchell Institute for Fundamental Physics \& Astronomy,\\\ \ Department of Physics \& Astronomy, Texas A\&M University, College Station, TX 77843, USA}
\author[0000-0002-1775-4859]{Lucas M.~Macri}
\affiliation{George P.~and Cynthia W.~Mitchell Institute for Fundamental Physics \& Astronomy,\\\ \ Department of Physics \& Astronomy, Texas A\&M University, College Station, TX 77843, USA}
\author{Shiyuan He}
\affiliation{Department of Statistics, Texas A\&M University, College Station, TX 77843, USA}
\author{Jianhua Z.~Huang}
\affiliation{Department of Statistics, Texas A\&M University, College Station, TX 77843, USA}
\author{Shashi M.~Kanbur}
\affiliation{Department of Physics, The State University of New York at Oswego, Oswego, NY 13126, USA}
\author[0000-0001-8771-7554]{Chow-Choong Ngeow}
\affiliation{Graduate Institute of Astronomy, National Central University, Jhongli 32001, Taiwan}
\correspondingauthor{Lucas M.~Macri}
\email{lmacri@tamu.edu}

\begin{abstract} 
We study the near-infrared properties of 690 Mira candidates in the central region of the Large Magellanic Cloud, based on time-series observations at $JHK_s$. We use densely-sampled $I$-band observations from the OGLE project to generate template light curves in the near infrared and derive robust mean magnitudes at those wavelengths. We obtain near-infrared Period-Luminosity relations for Oxygen-rich Miras with a scatter as low as 0.12~mag at $K_s$. We study the Period-Luminosity-Color relations and the color excesses of Carbon-rich Miras, which show evidence for a substantially different reddening law.
\end{abstract} 

\section{Introduction}

The accurate calibration of the Extragalactic Distance Scale plays a critically important role in understanding the content and evolution of the Universe. The most recent robust measurement of the Hubble constant ($H_0$) by \citet{2016ApJ...826...56R}, based on Cepheid variables and type Ia supernovae, and the inferred value of $H_0$ by \citet{2016A&A...594A..13P} under the assumption of $\Lambda$CDM are discrepant at the $3.4\sigma$ level. This may be an indication of additional components in the cosmological model, such as ``dark radiation'' \citep{2016JCAP...10..019B}. Since Mira variables (Miras) exhibit several promising properties as distance indicators, they may provide an independent route to $H_0$ to test the current discrepancy. Mira Period-Luminosity relations (PLRs) are relatively tight at near-infrared (NIR) wavelengths \citep{1981Natur.291..303G}, which are preferred over measurements in optical bands given the reduced effect of dust at longer wavelengths. The initial masses of Miras are considerably lower than those of Cepheids \citep[$1-2$ to $\lesssim 4 M_\sun$;][]{2009asrp.proc...48F}, which implies that Miras can be found in all types of galaxies, including those without recent star formation. Furthermore, Miras are generally more luminous than Cepheids, by $2-3$ magnitudes at $K_s$. A discussion of the importance of Miras as distance indicators can be found in \cite{2014EAS....67..263W}. In this article, we will use ``Miras'' to refer to both spectroscopically-confirmed Miras and Mira candidates (based only on photometric observations).

The PLRs of Mira-like variables have been studied by many generations of astronomers \citep{1928PNAS...14..963G,1929MeLuS..53....3G,1942ApJ....95..248W}. These initial studies were extended to filter-specific and bolometric PLRs once photometric systems were introduced. For example, \cite{1975ApJ...195..661E} related the periods and colors for the long-period variables and thus obtained their bolometric PLR; \cite{1981MNRAS.196..111R} determined the Mira PLRs in NIR and bolometric magnitudes. A sample of Miras in the LMC was studied by \cite{1982MNRAS.199..245G}, who found a $K$-band PLR dispersion of 0.23 mag. \cite{1987ASSL..132...51G} observed a larger sample of LMC Miras and obtained an $K$-band PLR for M-type Miras with scatter of only 0.13 mag. \cite{1989MNRAS.241..375F} noticed that several Oxygen-rich (O-rich) Miras with periods longer than 420 days are brighter than the extrapolated $K$-band PLR of a sample with shorter periods. The additional luminosity of these stars was explained by \cite{2003MNRAS.342...86W} as a consequence of an ongoing or a recently-experienced ``hot bottom burning'' episode \citep[HBB; see \S7.6 of][]{2013MNRAS.434..488M}. Excluding these very long-period stars, \cite{2008MNRAS.386..313W} concluded that O-rich and Carbon-rich (C-rich) Miras exhibit similar PLRs by studying a relatively small sample in the LMC.

The broader topic of PLRs for various types of luminous red giants (including but not limited to Miras) benefited immensely from the microlensing surveys towards the Magellanic Clouds. \citet{1999IAUS..191..151W} and \citet{2000PASA...17...18W} used observations from the MACHO project \citep{1993ASPC...43..291A} to establish the existence of five sequences in the Period-Luminosity plane, one of which corresponds to Miras pulsating in the fundamental mode. The OGLE project \citep{2008AcA....58...69U} enabled many significant refinements to these sequences \citep{2004AcA....54..129S,2004AcA....54..347S,2007AcA....57..201S}, which were recently examined from a theoretical perspective by \citet{2015MNRAS.448.3829W}.

In this article, we present sparsely-sampled NIR observations of over 600 LMC Miras originally discovered by the OGLE project \citep{2009AcA....59..239S}. Using their exquisite $I$-band photometry, we develop light curve templates for individual Miras at $JHKs$ and use them to derive PLRs with scatter as low as 0.12~mag. Miras are known to lose considerable amounts of mass during their evolution in the AGB phase \citep{1990ASPC...11..355W}. That material forms dust envelopes that attenuate starlight to varying degrees and may hinder accurate Mira-based distance determinations. We compare the color indices of O- and C-rich subtypes of Miras in our sample and find that they have significantly different properties, as already known from previous work \citep[e.g.,][]{1982MNRAS.199..245G}. The color indices of O-rich Miras exhibit much less dispersion than those of C-rich ones, indicating very different circumstellar dust environments.

The rest of the article is organized as follows. In Section~\ref{sec:data} we describe the data used in this study. The procedure to generate NIR light curve templates is described in Section~\ref{sec:tpl}. Section~\ref{sec:color} presents an analysis of the NIR color excesses, while Section~\ref{sec:plr} presents Period-Luminosity and Period-Luminosity-Color relations for O- and C-rich Miras, respectively. We draw conclusions and discuss the results in Section~\ref{sec:disc}.

\section{Data} \label{sec:data}

This study used $I$-band measurements of Miras by OGLE-\rom{3} \citep{2009AcA....59..239S} and $JHK_s$ observations from the LMC Near-Infrared Synoptic Survey \citep[LMCNISS;][]{2015AJ....149..117M}.

\subsection{LMCNISS Measurements}

LMCNISS observed the central region of the LMC in $JHK_s$ using the CPAPIR camera at the CTIO 1.5m telescope. A detailed description of the survey was presented by \citep{2015AJ....149..117M}. The data products from this survey include observations for 690 Miras. 681, 679, and 676 variables have measurements in $JHK_s$, respectively, with 668 observed in all three bands. A total of 84,852 individual photometric measurements are available, which can be separated into three broad groups based on the observation dates: 2006 November, 2007 January 2007 November. Compared to the periods of Miras, the time span of any of the groups is short enough to reject extreme outliers based on the mean magnitude of a given variable within the group. Across this study, ``extreme outliers'' are defined as data beyond outer fences, which are mathematically equal to $Q_1 -3\times(Q_3 -Q_1)$ and $Q_3 +3\times(Q_3 -Q_1)$ with $Q_1$ and $Q_3$ being the values at the 25 and 75 percentiles, respectively. We rejected such extreme outliers, which constituted 3.6\% of the entire data. The final median number of measurements per Mira is 42, 44, and 38 in $JHK_s$, respectively. 

\subsection{OGLE-\rom{3} Measurements \label{sec:iband}}

We retrieved the $I$-band light curves and the O/C-rich classification for each Mira observed by LMCNISS from the OGLE-\rom{3} database. The OGLE-\rom{3} light curves are densely sampled and have excellent signal-to-noise levels. The observations extend for as long as 12 years for most Miras and gracefully cover the period when the LMCNISS was carried out. We used 551,923 measurements and rejected only 46 data points that noticeably deviated from their neighbors, possibly due to occasional photometry artifacts.

We used the semi-parametric Gaussian process model of \citet{2016AJ....152..164H} to fit the OGLE-\rom{3} light curves. This model assumes that the $I$-band light curve of a Mira can be decomposed into a sinusoidal component with a single period $P$ and an amplitude $A(I)$, plus a data-driven component that captures cycle-to-cycle and long-term variations. This differs from the approach taken by the OGLE-\rom{3} team, which solved for multiple periods for each variable. We assumed that the period and amplitude of the sinusoidal component of a given Mira do not change during the time window being considered. To avoid large uncertainties in some regions of template curves where data are sparse, we excluded any sampling gaps longer than 50 days. As a result, the template curves are not continuous curves. Thanks to the dense sampling of the $I$-band light curve and the exclusion of large gaps from the analysis, the choice of fitting method (Gaussian process, smooth spline or piecewise high-order polynomial) does not impact the result.

\section{Template Mira Light Curves in the NIR} \label{sec:tpl}

In order to study the NIR PLRs, we estimated mean $JHK_s$ magnitudes with the help of template light curves at these wavelengths. The method we used is slightly complicated due to several restrictions. Firstly, unlike Cepheids or RR Lyrae, Mira light curves are not strictly repetitive and phase-folding does not yield smooth curves. Secondly, given the long periods of Miras, the LMCNISS observations are extremely sparse and are concentrated at only three ``epochs,'' which can by no means be used to estimate the true underlying light curves without additional information. Lastly, the amplitudes of Mira light curves are known to be large and vary from cycle to cycle. Simply averaging these three ``epochs''  would result in unacceptably large scatter in the PLRs. 

To overcome the aforementioned restrictions, we made use of the densely-sampled $I$-band light curve data from OGLE-\rom{3}, which span a much longer baseline that includes the times of the $JHK_s$ observations. We assumed that as long as the $JHK_s$ observations take place within the time window sampled by the $I$-band light curves, these will track each other. We use the $J$ band to illustrate our method. The $H$- and $K_s$-band light curves were derived using the same approach.

\begin{figure}
\epsscale{1}
\plotone{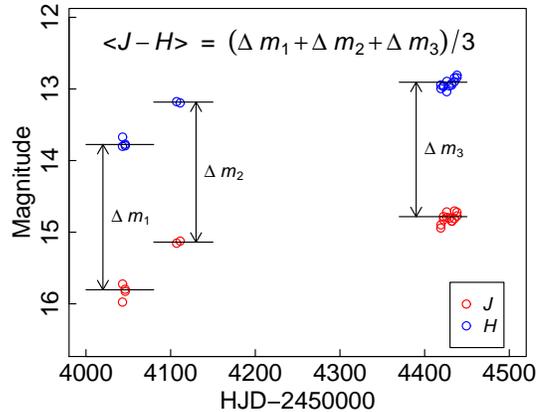}
\caption{Example of how the mean value of $J\,-\,H$ (used in the linear regression model) was derived from the three groups of measurements for OGLE-LMC-LPV-08476.\label{fig_ccol}}
\end{figure}

\begin{figure*}
\epsscale{0.99}
\plotone{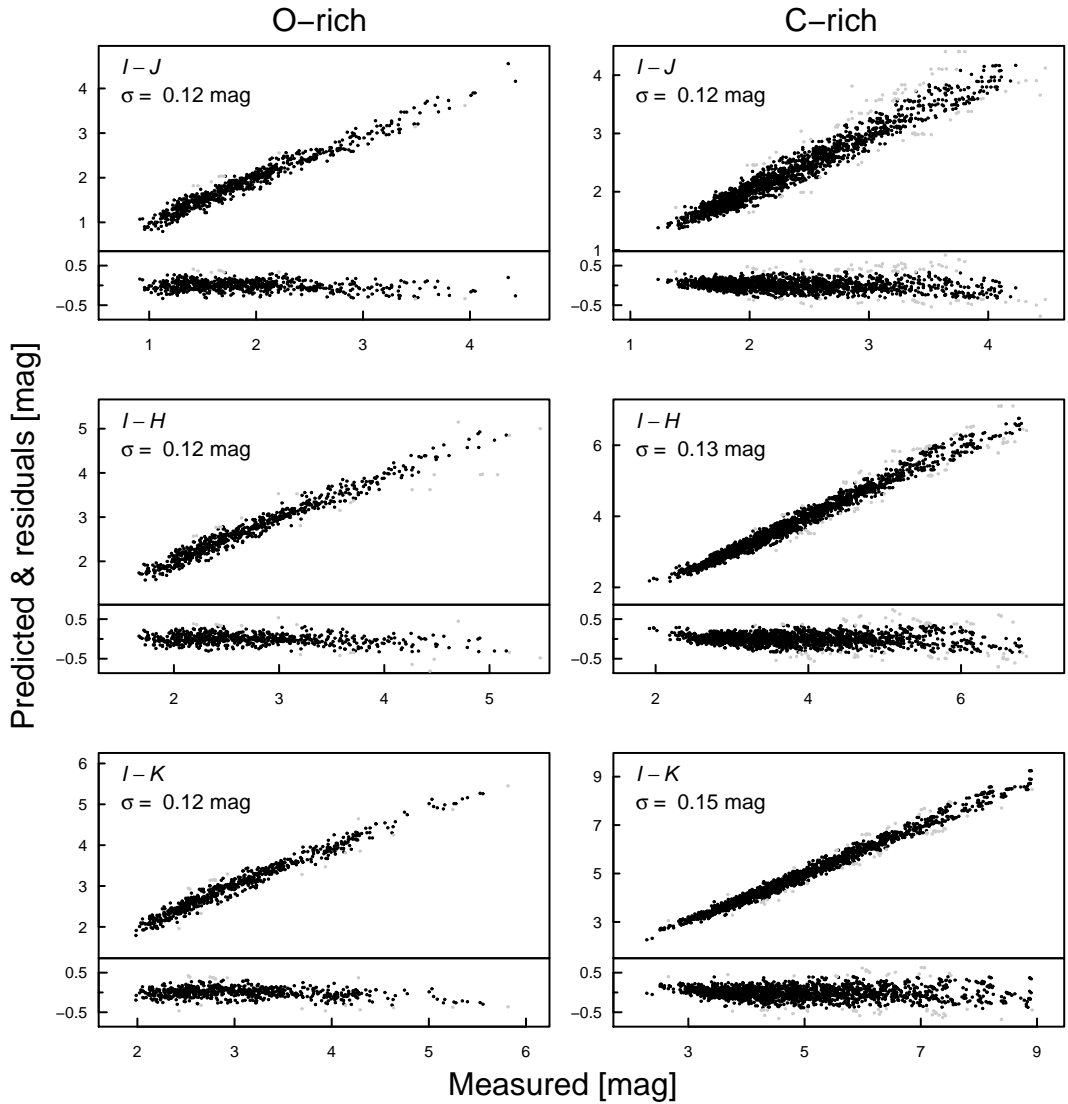}
\caption{Results of the two-stage least-square regression to predict $I\,-\,JHK_s$ colors for O- and C-rich Miras.\label{regression}}
\end{figure*}

We firstly built a regression model to predict the value of $I(t)\,-\,J(t)$ (where $t$ is the time of observation). We then used the relation provided by the regression model and the $I$-band light curve to derive the $J$-band light curve for each Mira, i.e.
\begin{equation}
J(t) = I(t)- \big(I(t)-J(t)\big),\nonumber
\end{equation}
where $I(t)$ is obtained by the process described in \S\ref{sec:iband} and $I(t)\,-\,J(t)$ is provided by the regression model.

We built a regression model with $(I(t)\,-\,J(t))$ as the response using several variables as possible predictors: phase $\phi(t)$, absolute $I$ magnitude $I_\mathrm{abs}(t)$, period $P$, amplitude of the sinusoidal component of the $I$-band model $A(I)$, and mean NIR color indices $(J-H)$ and $(H-K)$. The phase is defined relative to the time of maximum light in the $I$ band and is restricted to the interval $0\,\leq\,\phi\,<\,1$. The mean NIR color indices were calculated as shown in Figure~\ref{fig_ccol}: we first computed their values for each temporal group, and then obtained the average using equal weights for each point. We considered several functional forms for each variable that could contribute to the predictor, which are listed in Table~\ref{tbl_lmp}.

We built separate regression models for each of $I\,-\,JHK_s$, and treated O and C-rich subtypes separately.  For each of these six groups, we first selected a subset of the predictors using the LASSO algorithm \citep{Tibshirani1994} to avoid over-fitting. The regularization parameter was determined by a 10-fold cross-validation. With the selected subset of predictors, we trained a least-square regression with two stages. In the first stage, we ran the least-square regression using all data, while in the second stage, we removed residual outliers in the first stage and reran the same least-square regression. Figure~\ref{regression} shows the result of this two-stage procedure, and the regression coefficients for each of the six groups are presented in Table~\ref{tbl_lmp}. The typical scatter of residuals is 0.12~mag.

With these regression models, we predicted the color curves and hence obtained the $J(t)$, $H(t)$, and $K_s(t)$ curves. For each Mira in each band, we then solved an overall offset between the predicted curve and actual measurements. The final NIR template curves were obtained by subtracting the offsets from the predicted curves.

\begin{center}
\begin{deluxetable*}{lrrrrrr}
\tablecaption{Predictors and Coefficients for the Regression Model\label{tbl_lmp}}
\tabletypesize{\normalsize}
\tablewidth{\textwidth}
\tablehead{
\colhead{Predictor}  & \multicolumn6c{Regression Coefficients} \\ \cline{2-7}
& \multicolumn3c{O-rich} & \multicolumn3c{C-rich} \\
 \multicolumn1r{$(I - )$} & \colhead{$J$} & \colhead{$H$} & \colhead{$K_s$} & \colhead{$J$} & \colhead{$H$} & \colhead{$K_s$}
}
\startdata
              $\ast I_\mathrm{abs}$ &  7.9e-01 &  7.4e-01 &  7.3e-01 &  3.4e-01 &  3.7e-01 &  4.4e-01\\
            $\ast I_\mathrm{abs}^2$ &       NA &       NA &       NA & -2.0e-03 & -1.1e-02 & -3.5e-03\\
 $\ast F = 10^{-0.4I_\mathrm{abs}}$ &  2.1e-03 &       NA &       NA &       NA &       NA & -4.8e-04\\
                         $\ast F^2$ &       NA &  1.4e-05 &  8.9e-06 &  2.8e-06 &       NA &       NA\\
                         $\ast F^3$ & -1.2e-08 & -3.7e-08 & -2.0e-08 &       NA &       NA &       NA\\
                     $\ast F^{0.5}$ &       NA &       NA &       NA &       NA &       NA & -4.5e-02\\
              $\ast \cos(2\pi\phi)$ &       NA & -8.4e-03 & -4.0e-02 &  6.1e-02 &  4.6e-02 &  4.3e-02\\
              $\ast \sin(2\pi\phi)$ &  1.9e-01 &  2.0e-01 &  2.1e-01 &  7.2e-02 &  8.3e-02 &  7.7e-02\\
              $\ast \cos(4\pi\phi)$ &  1.4e-02 &  1.1e-02 &  1.5e-02 & -3.5e-03 & -1.9e-03 &       NA\\
              $\ast \sin(4\pi\phi)$ &  9.4e-03 &  1.6e-02 &  1.8e-02 &  1.7e-02 &  1.8e-02 &  1.8e-02\\
              $\ast \cos(6\pi\phi)$ &  1.7e-02 &  7.7e-03 &  7.7e-03 &  5.2e-03 &  6.6e-03 &       NA\\
              $\ast \sin(6\pi\phi)$ &       NA &       NA &       NA & -4.8e-03 & -9.0e-03 & -1.3e-02\\
                                $P$ & -7.7e-02 &  1.9e-02 &  3.4e-02 &  2.5e-03 &  5.4e-03 &  5.4e-03\\
                           $\log P$ &  6.3e+03 & -3.2e+03 & -5.5e+03 &       NA & -2.6e+01 & -3.3e+01\\
                          $\log^2P$ & -1.1e+03 &  4.8e+02 &  8.4e+02 &       NA &       NA &       NA\\
                          $\log^3P$ &  1.1e+02 & -4.5e+01 & -7.9e+01 &       NA &       NA &       NA\\
                      $\log^{0.5}P$ & -9.5e+03 &  5.0e+03 &  8.7e+03 &       NA &  7.7e+01 &  1.0e+02\\
                              $J-H$ & -5.6e-01 &  3.9e-01 &  3.5e-01 & -3.5e-01 &  4.8e-01 &  3.2e-01\\
                              $H-K$ & -5.4e-01 & -4.8e-01 &  4.1e-01 &       NA & -2.4e-01 &  3.9e-01\\
                              $A_I$ &  7.9e-03 &  1.7e-02 &  1.7e-02 &  5.0e-02 &  7.4e-02 &  1.0e-01\\
                          Intercept &  4.2e+03 & -2.3e+03 & -4.0e+03 &  2.7e+00 & -5.4e+01 & -7.3e+01\\
\hline
                     $\sigma$ [mag] &     0.12 &     0.12 &     0.12 &     0.12 &     0.13 &     0.15\\
\enddata
\tablecomments{Predictors beginning with $\ast$ indicates time-dependent quantities.\\ \ \ NA indicates predictors that were rejected by the LASSO regularization.}
\end{deluxetable*}
\begin{figure*}[!t]
\epsscale{1.2}
\plotone{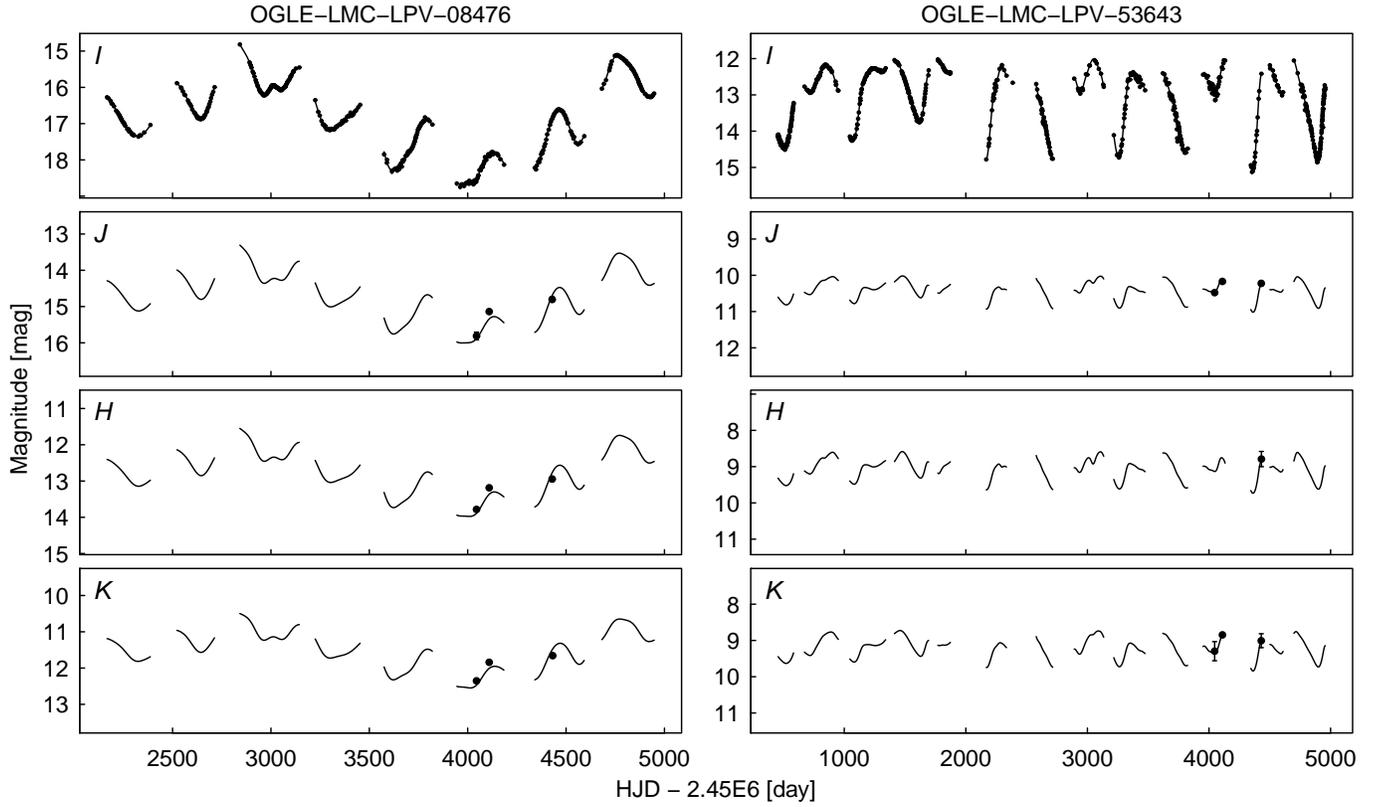}
\caption{Template light curves in $IJHK_s$ for Miras OGLE-LMC-LPV-08476 (left) and -53643 (right). The measurements are indicated by black points. \label{fig_flc}}
\end{figure*}

\begin{center}
\begin{deluxetable*}{cccrrrrrrrrrrrrcc}[!h]
\tabletypesize{\scriptsize}
\tablecaption{$JHK_s$ Magnitudes for LMC Miras\label{tbl_mag}}
\tablewidth{0pt}
\tabletypesize{\footnotesize}
\tablehead{
\multicolumn{1}{c}{OGLE ID} & \multicolumn{1}{c}{flag$^a$} & St$^b$ & \multicolumn{3}{c}{$J$ (mag)} & $\sigma$ & \multicolumn{3}{c}{$H$ (mag)} & $\sigma$ & \multicolumn{3}{c}{$K$ (mag)} & $\sigma$ & Period & $t_\mathrm{ref}^c$\\
& & & mean & max &  min  && mean & max & min && mean & max & min && \multicolumn2c{(days)}}
\startdata
08476 & 0 & C & 14.72 & 16.00 & 13.31 & 0.11 & 12.79 & 13.97 & 11.55 & 0.12 & 11.52 & 12.55 & 10.50 & 0.12 &  324.4 & 3466.7 \\
08922 & 0 & C & 14.17 & 14.74 & 13.42 & 0.03 & 12.52 & 13.03 & 11.87 & 0.03 & 11.26 & 11.69 & 10.71 & 0.05 &  375.2 & 3566.9 \\
09071 & 0 & C & 13.10 & 13.68 & 12.26 & 0.01 & 11.52 & 12.02 & 10.80 & 0.03 & 10.37 & 10.77 &  9.81 & 0.06 &  463.0 & 3681.0 \\
\enddata
\tablecomments{Table~\ref{tbl_mag} is published in its entirety in machine-readable format. The first 3 (out of 690 lines) are shown here for guidance regarding its form and content. (a): $0$, magnitude from regression model; $c$, outlier in color-color diagram; $r$, regression outlier; $m$, missing $J-H$ or $H-K$ measurement; $p$, phase may be incorrect. (b): Mira subtype (O- or C-rich). (c): HJD - 2450000.}
\end{deluxetable*}
\end{center}
\end{center}

\clearpage
Figure~\ref{fig_flc} shows an example of NIR template curves for two Miras, including a long-period object that exhibits double maxima. Our modeling procedure is able to accommodate these types of pulsators without any issues. We obtained the mean, maximum, and minimum $JHK_s$ magnitudes of each variable by evaluating the simple average, maximum, and minimum of the piecewise template curves. These values are listed in Table~\ref{tbl_mag}.

We applied the above analysis to 658 of the 690 Miras in our sample. The other variables were excluded because they lacked NIR color indices (22), had highly unusual values for these quantities (6), exhibited unrealistic spikes in the template curves perhaps due to incorrect phase determination (3) or had very large residuals (1). For these 32 Miras we simply computed mean $JHK_s$ magnitudes by taking the average of the mean measurements of individual temporal groups. The mean magnitudes for these Miras are also listed in Table~\ref{tbl_mag} and were included in the subsequent analysis.

\section{Color Excesses of Miras} \label{sec:color}
\begin{figure}
\epsscale{1}
\plotone{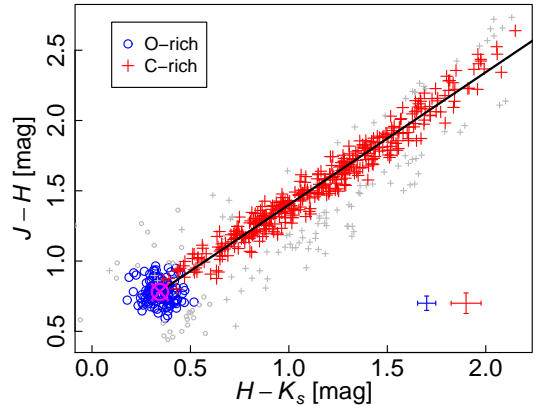}
\caption{Observed color-color diagram for LMC Miras. The O- and C-rich subtypes are indicated by blue circles and red pluses, respectively. Outliers are indicated by grey symbols. The magenta circled cross shows the centroid of O-rich Mira color indices. The black line is the first-order fit for the C-rich Mira colors. \label{fig_ccd}}
\end{figure}

The O- and C-rich subtypes of Miras are known to exhibit different NIR colors \citep{1982MNRAS.199..245G,1982MNRAS.201..439F,1983ApJ...272...99W,2009AcA....59..239S}, due to their different circumstellar dust environments \citep{2011MNRAS.412.2345I,2013MNRAS.434.2390N,2015A&A...575A.105B,2016A&A...594A.108H,2017ApJ...835...77M}. We studied the circumstellar extinction of the LMC Miras in our sample using the $J-H$ and $H-K_s$ color indices. We found that the LMC Miras clustered in two regions in the observed color-color diagram as shown in Figure~\ref{fig_ccd}. Most O-rich Miras (blue circles) are centered at [0.345$_{\pm 0.005}$, 0.780$_{\pm 0.006}$], although we note that our sample does not contain some previously-known O-rich LMC Miras that are very red \citep{1998A&A...329..169V,2003MNRAS.342...86W}. The C-rich Miras (red pluses) are located along a narrow strip. We obtained the centroid of the O-rich Mira cluster, excluding points outside of a 0.2 mag radius. We fit a straight line to the C-rich Miras iteratively, with the first iteration excluding any points further than 0.2 mag from the fit, and the second iteration excluding extreme outliers. Interestingly, the fitted line 
\begin{displaymath}
(J-H) - 0.780 = 0.943_{\pm 0.010} [(H-K) - 0.345]
\end{displaymath}
passes through the centroid of O-rich Mira cluster well within its uncertainty. This suggests both Mira subtypes may have similar intrinsic colors in the absence of dust.

\begin{figure}
\epsscale{1}
\plotone{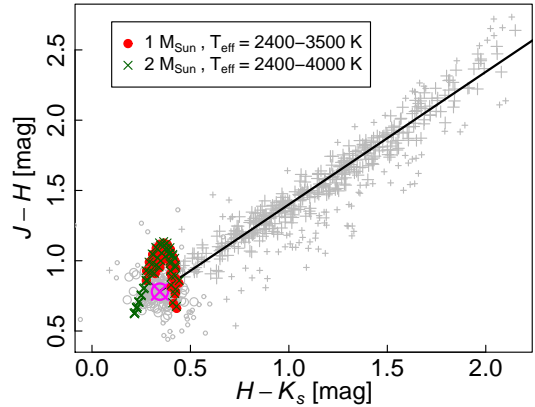}
\caption{Same as Figure~\ref{fig_ccd} but showing the intrinsic color indices of C-rich giants from hydrostatic models \citep{2009A&A...503..913A}. Red filled circles show $M = 1 M_\sun$ models while the green crosses indicate $M = 2 M_\sun$ models. The observed color indices of C-rich and O-rich Miras are shown in gray pluses and circles, respectively.\label{fig_aringer}}
\end{figure}
 
\begin{figure}
\epsscale{1}
\plotone{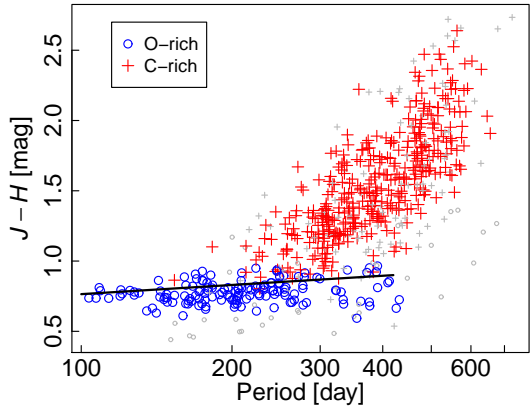}
\caption{Correlation between $J-H$ color index and period for C- and O-rich Miras (red pluses and blue circles, respectively). The black line indicates the relation derived by \citet{1995MNRAS.273..383G} using observations of O-rich Miras in the LMC by \citet{1989MNRAS.241..375F}. Their relation was placed in the 2MASS photometric system using the transformations from \citet{2001AJ....121.2851C}.\label{fig_clrP}}
\end{figure}

To confirm the intrinsic colors of C-rich Miras, we obtained model-based color indices. In lieu of Mira pulsation models, we used the C-rich giant hydrostatic model by \cite{2009A&A...503..913A}, which provide synthetic photometry and luminosities in various filters, including 2MASS $JHK_s$. We restricted their results to those with $\log(g) = 0$, $Z = Z_\sun$, and $M = 1 M_\sun$ or $2M_\sun$. The model colors are shown in Figure~\ref{fig_aringer}. We found that the model colors for C-rich giants are consistent with the observed colors of O-rich Miras, which supports the assumption that two subtypes of Miras have similar intrinsic color indices, while only the C-rich Miras are surrounded by a considerable amount of circumstellar dust that noticeably reddens the light. Our observations also reproduce the previously-established relation \citep{1989MNRAS.241..375F,1995MNRAS.273..383G} that the observed color indices for both subtypes of Miras increase as a function of period, as shown in Figure~\ref{fig_clrP}.

\begin{figure}
\epsscale{1}
\plotone{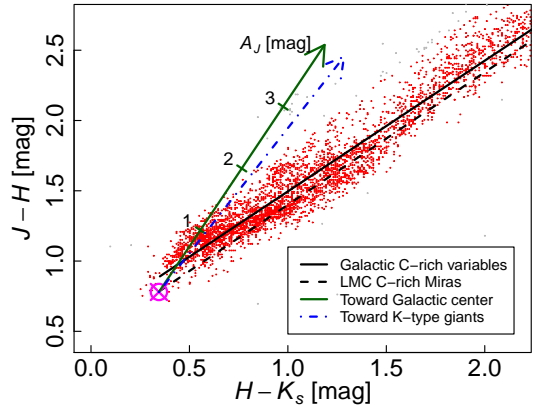}
\caption{Observed color indices for Galactic C-rich variable stars (red dots) from \cite{2006MNRAS.369..751W} and best-fit linear relation (black solid line). The green solid and blue dash-dotted lines indicate the reddening vectors towards the Galactic Center \citep{2009ApJ...696.1407N} and towards Galactic K giants \citep{2014ApJ...788L..12W}, respectively. The black dashed line indicates the best-fit linear relation for C-rich Miras in the LMC.\label{fig_whitelock}}
\end{figure}

We compared our color measurements to those of Galactic C-rich variable stars \citep{2006MNRAS.369..751W}, as well as the extinction laws of interstellar dust. With the assumption that O- and C-rich Miras have the same $J-H$ and $H-K_s$ color indices, the color excess ratio $E(J-H)/E(H-K_s)$ for C-rich Miras simply equals the slope in the color-color diagram, which is $0.94\pm0.01$. We transformed the photometric measurements of \cite{2006MNRAS.369..751W} to 2MASS $JHK_s$ magnitudes with Equations 33--36 of \cite{2001AJ....121.2851C} and obtained a similar color excess ratio ($0.930_{\pm0.003}$), as shown in Figure~\ref{fig_whitelock}. This value, however, is very different from the interstellar extinction laws. For example, \cite{2009ApJ...696.1407N} found $E(J-H)/E(H-K_s) = 2.09$ towards the direction of Galactic center; \cite{2014ApJ...788L..12W} derived a value of 1.78 using a sample of 5942 K-type giants. This difference is not surprising given the wide range of sources that contribute to interstellar dust \citep[see][and references therein]{2009MNRAS.396..918M}.

\section{Mira PLRs in the NIR}\label{sec:plr}

Miras are not strictly periodic but exhibit chaotic variations in their light curves. Given the limited time coverage of the observations, it is not possible to obtain their true mean magnitudes. As a proxy, we used the median value of all the maxima and minima in each piecewise template light curve. For those objects without template light curves, we used the average magnitude of the three temporal groups. We subtracted the distance modulus of the LMC \citep[18.493~mag;][]{2013Natur.495...76P} to obtain their absolute values. The uncertainties of the $JHK_s$ photometric zeropoints are $\sim 0.02$~mag and we caution the readers that these zeropoint errors, together with the uncertainty in the LMC distance modulus, were not included in the subsequent analysis.

\begin{figure*}
\epsscale{1.2}
\plotone{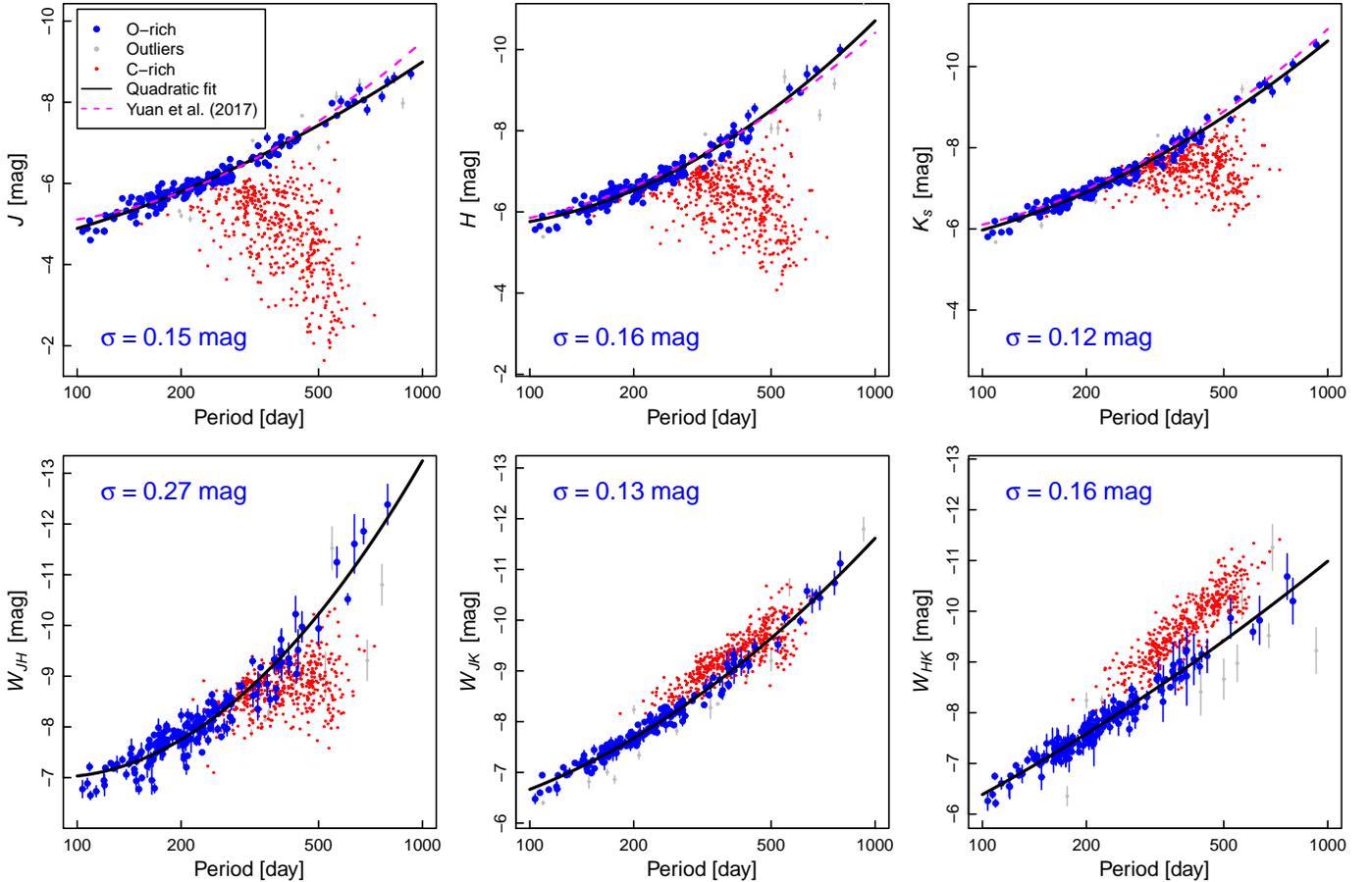}
\caption{PLRs of the LMC Miras. The blue points indicate O-rich Miras while the red points indicates C-rich Miras. The black solid lines show the equally weighted quadratic fit to the O-rich Mira PLRs, while the magenta dashed lines show the PLRs based on single-epoch 2MASS observations \citep{2017AJ....153..170Y}. \label{fig_lmcplr}}
\end{figure*}

We fit empirical quadratic relations
\begin{equation}
M = a_0 + a_1(\log P - 2.3) + a_2(\log P - 2.3)^2
\end{equation}
to the periods and magnitudes of O-rich Miras. We rejected extreme outliers in each band during the fit. The O-rich PLRs derived in this work are broadly consistent with the preliminary relations derived by \cite{2017AJ....153..170Y} based on single-epoch 2MASS observations. However, our relations exhibit $\sim 2\times$ lower dispersion. The relations are plotted in Figure~\ref{fig_lmcplr} and summarized in Table~\ref{tbl_lmcplr}.

We also tested the PLRs using Wesenheit indices based on the interstellar extinction law from \citet{2007A&A...476...73F}, though this may be different from the circumstellar dust extinction law for Miras. These Wesenheit indices are 
\begin{equation}
  \begin{aligned}
    & W_{JH} = H - 1.611\cdot(J-H) \\ &  W_{JK} = K_s - 0.679\cdot(J-K_s)  \\ & W_{HK} = K_s - 1.900\cdot(H-K_s)
  \end{aligned}
\end{equation}
 We found that the $K_s$-band PLR gives the least scatter in all six bands or Wesenheit indices. We caution the readers that these Wesenheit indices may be inappropriate as distance indicators since the circumstellar dust extinction law and the color dependences of PLRs for Miras are not fully understood at present.

The recent analysis by \citet{2015MNRAS.448.3829W} gives a strong motivation to consider a linear formulation of these relations for O-rich Miras, as long as one excludes variables with $P>400$~d which may be affected by HBB (as discussed in the Introduction). We list the results of these fits in Table~\ref{tbl_lmcplr} and note that as long as the aforementioned upper period limit is imposed, these relations exhibit comparable scatter to the quadratic formulations that span the entire period range.

\begin{figure*}
\epsscale{0.9}
\plotone{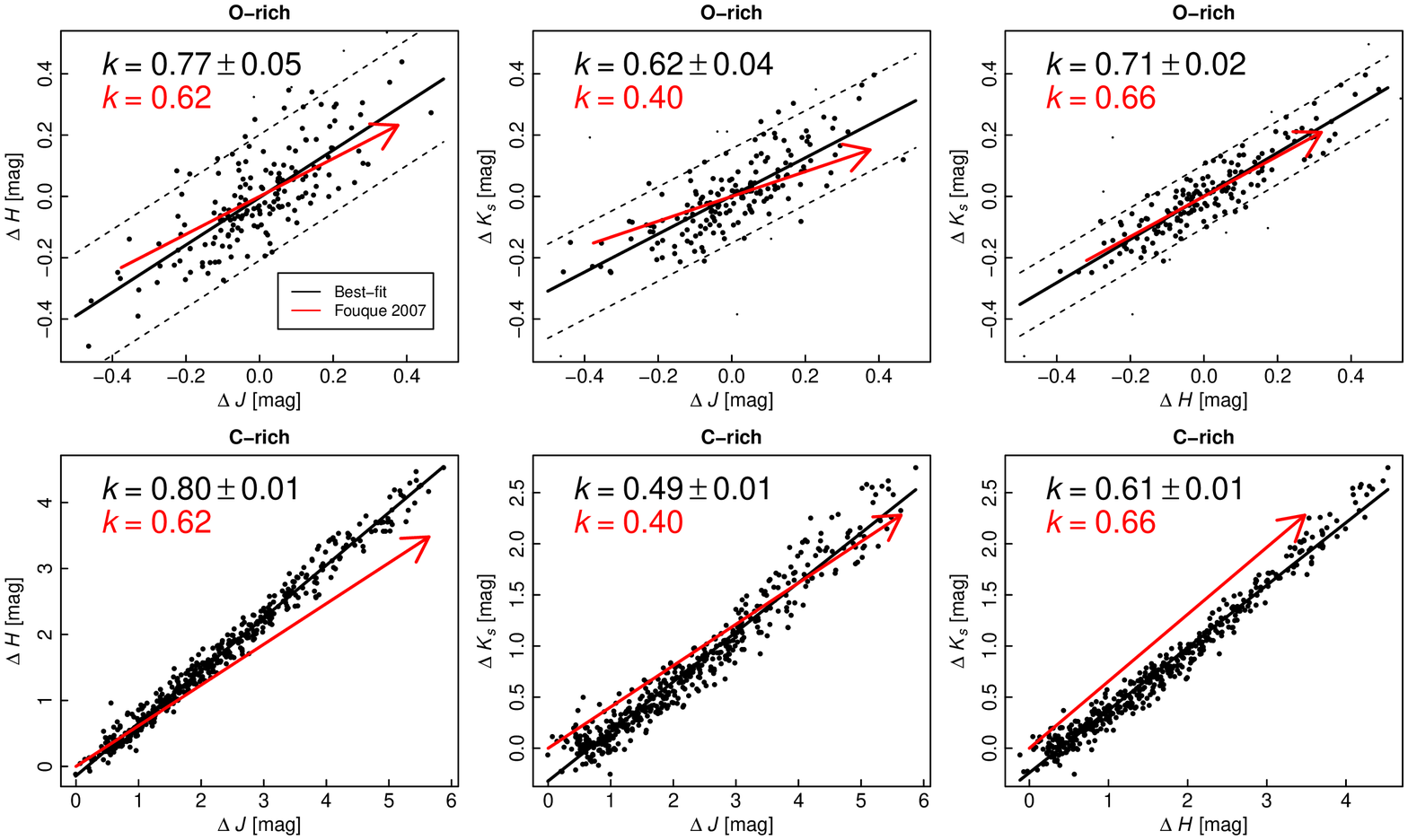}
\caption{Correlations of PLR residuals between different bands for O-rich Miras (upper panels) and C-rich Miras (lower panels) in the LMC. We adopted the PLRs of O-rich Miras to compute the ``residuals'' for C-rich Miras. The black solid lines indicate the first-order fit to correlations of residuals, while the black dashed lines indicate the $\pm 2\sigma$ widths of the relations. The red arrows indicate the direction of interstellar reddening based on \citet{2007A&A...476...73F}.  \label{fig_lmcrescor}}
\end{figure*}

We explored the correlations of the PLR residuals across different bands, defined as $M_\mathrm{obs} - M_\mathrm{PLR}$. For the C-rich Miras, we used the PLRs of O-rich Miras as fiducial relations to compute ``residuals''. The same method has been used by \citet{2011MNRAS.412.2345I} to investigate the circumstellar extinction of Miras. We found strong correlations of the residuals across $JHK_s$ bands for both O- and C-rich Miras, as shown in Figure~\ref{fig_lmcrescor}. Several factors may contribute to these correlations: (1) color variations due to changes in temperature \citep[see the related discussion in][]{1991PASP..103..933M}; (2) interstellar dust; (3) circumstellar dust; (4) correlated measurement noise. For C-rich Miras, the correlations span ranges that are too large to be explained by factors (2) and (4). Nevertheless, the validity of using PLRs of O-rich Miras as fiducials for this purpose remains to be proven. We compared the slopes of these correlations to the theoretical interstellar reddening vectors derived from \citet{2007A&A...476...73F} (red arrows in Figure~\ref{fig_lmcrescor}) and found they are statistically different in most cases. 
\begin{figure*}[!hbtp]
\epsscale{0.9}
\plotone{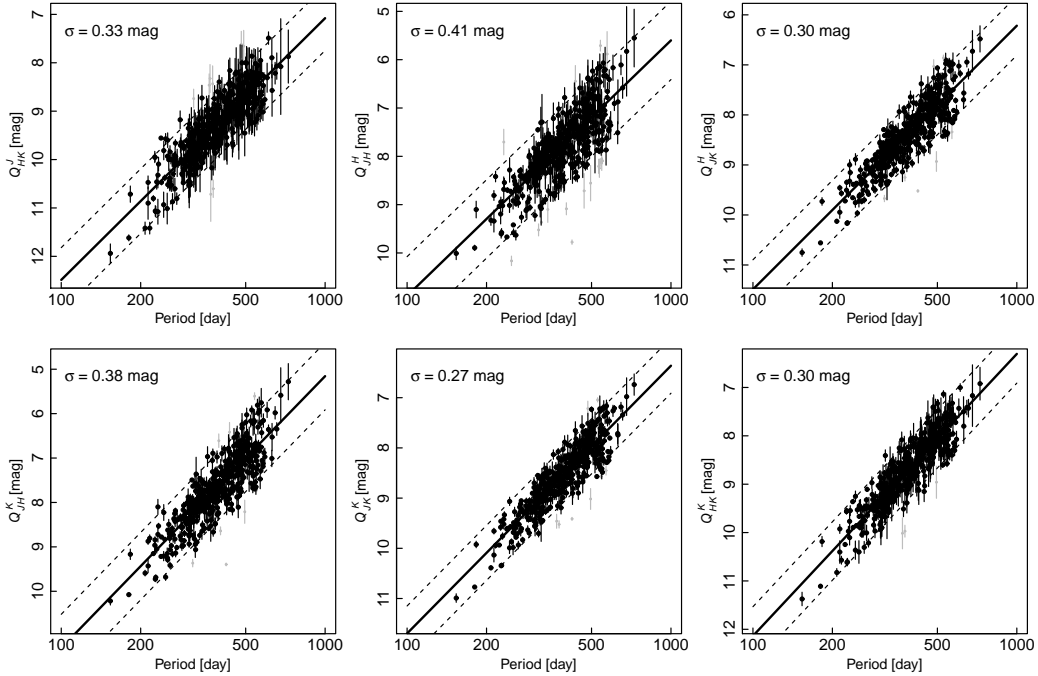}
\caption{The oPLCRs of C-rich Miras in various combinations of bands and color terms in the LMC. The black solid lines indicate the first-order fits to the oPLCRs, while the black dashed lines indicate the $\pm 2\sigma$ widths of the relations. Extreme outliers are indicated by gray points.\label{fig_lmcoPLCRs}}
\end{figure*}
\begin{deluxetable}{lrrrrrr}
\tablecaption{NIR PLRs for O-rich Miras in the LMC\label{tbl_lmcplr}}
\tabletypesize{\footnotesize}
\tablewidth{0pt}
\tablehead{\colhead{Band} & \colhead{$a_0$} & \colhead{$a_1$} & \colhead{$a_2$} & \colhead{$\sigma$} & \colhead{$N_i$}  & \colhead{$N_f$}}
\startdata
\multicolumn{7}{l}{Quadratic relations}\\
\cline{1-7}
$J$     & -5.80 (1) & -3.49 (09) & -1.54 (23) & 0.15 & 187 & 178\\
$H$     & -6.53 (1) & -3.59 (10) & -3.40 (31) & 0.16 & 183 & 173\\
$K_s$   & -6.90 (1) & -3.77 (08) & -2.23 (20) & 0.12 & 183 & 176\\
$W_{JH}$& -7.75 (2) & -4.02 (16) & -5.47 (52) & 0.27 & 181 & 171\\
$W_{JK}$& -7.67 (1) & -4.04 (08) & -2.28 (22) & 0.13 & 181 & 167\\
$W_{HK}$& -7.58 (1) & -4.25 (10) & -0.88 (30) & 0.16 & 182 & 171\\
\cline{1-7}
\multicolumn{7}{l}{{Linear relations ($P<400$~d)}}\\
\cline{1-7}
$J$      & -5.82 (1) & -3.48 (09) &\multicolumn{1}{c}{$\dots$}& 0.15 & 163 & 158\\
$H$      & -6.58 (1) & -3.64 (09) &\multicolumn{1}{c}{$\dots$}& 0.16 & 165 & 163\\
$K_s$    & -6.93 (1) & -3.77 (07) &\multicolumn{1}{c}{$\dots$}& 0.12 & 165 & 158\\
$W_{JH}$ & -7.83 (2) & -3.85 (15) &\multicolumn{1}{c}{$\dots$}& 0.25 & 163 & 158\\
$W_{JK}$ & -7.70 (1) & -4.05 (07) &\multicolumn{1}{c}{$\dots$}& 0.12 & 163 & 151\\
$W_{HK}$ & -7.60 (1) & -4.32 (09) &\multicolumn{1}{c}{$\dots$}& 0.15 & 165 & 160\\
\enddata
\tablecomments{$\sigma$: dispersion (mag). $N_i$: initial number of variables. $N_f$: final number after rejecting extreme outliers.}
\end{deluxetable}

\begin{deluxetable}{lrrrrrr}
\tablecaption{NIR oPLCRs for C-rich Miras in the LMC\label{tbl_oPLCRs}}
\tablewidth{0pt}
\tablehead{\colhead{Index} & \colhead{$b$} &\colhead{$a_0$} & \colhead{$a_1$} & \colhead{$\sigma$} & \colhead{$N_i$} & \colhead{$N_f$}}
\startdata
$Q^J_{HK}$ & 3.78 & 10.86 (4) & -5.41 (14) & 0.33 & 482 & 474 \\
$Q^H_{JH}$ & 2.89 &  9.30 (5) & -5.29 (17) & 0.41 & 482 & 465 \\
$Q^H_{JK}$ & 1.41 &  9.91 (4) & -5.27 (12) & 0.30 & 482 & 473 \\
$Q^K_{JH}$ & 2.20 &  9.44 (5) & -6.12 (16) & 0.38 & 482 & 474 \\
$Q^K_{JK}$ & 0.91 & 10.10 (4) & -5.33 (11) & 0.27 & 482 & 474 \\
$Q^K_{HK}$ & 1.98 & 10.39 (4) & -5.83 (13) & 0.30 & 482 & 479 \\
\enddata
\tablecomments{$\sigma$: dispersion (mag). $N_i$: initial number of
  variables. $N_f$: final number after rejecting extreme outliers.}
\end{deluxetable}

While we cannot break the degeneracy of these many factors, we investigated observed Period-Luminosity-Color Relations (oPLCRs) for C-rich Miras given the tight correlation of residuals. As we have mentioned, in terms of studying the circumstellar dust extinction and color dependences of PLRs for C-rich Miras, the interstellar extinction and correlated measurement errors are negligible. We fit the oPLCRs in the form of 
\begin{equation}
  Q_{JK}^K \equiv K_s - b\cdot (J-K_s) =  a_0 + a_1(\log P - 2.3)
\end{equation}
where $a_0$, $a_1$ and $b$ are free parameters. We applied a revised version of least-squares fit to minimize $\sigma\cdot[(1+b)^2\sigma_K^2 + b^2\sigma_J^2]^{-1/2}$ instead of $\sigma$ to account for the dependence of $\sigma$ on $b$, where $\sigma$ is scatter of fit to $Q_{JK}^K$. We used equal values of $\sigma_J$ and $\sigma_K$ since the intrinsic scatter is unknown. We found that the choice of the values of $\sigma_J$ and $\sigma_K$ or the number of orders in the period terms does not significantly change the results. For example, using the scatter ratio of PLRs of O-rich Miras ($\sigma_J/\sigma_K = 1.25$) or adding a quadratic term in the period only changes the value of $b$ by 2.6\% and 1.1\%, respectively. We excluded any measurements with uncertainties greater than 0.3 mag or extreme outliers during the fit. We also obtained the oPLCRs for other combinations of colors, and list the results in Table~\ref{tbl_oPLCRs}. We show the first-order fit to the oPLCRs in Figure~\ref{fig_lmcoPLCRs}. We are aware that the circumstellar dust extinction law may be degenerate with intrinsic color terms, and one should not take the $Q$ values as reddening-free Wesenheit indices. 

\section{Conclusions and Discussion} \label{sec:disc}

We analyzed the NIR color indices for over 600 LMC Miras. We found that the O- and C-rich subtypes occupy different yet well-confined regions in the color-color diagram. We compared our results with stellar model predictions and literature measurements of Galactic C-rich stars. We hypothesize that both Mira subtypes have similar intrinsic colors but only C-rich Miras exhibit significant attenuation by  circumstellar dust. If our hypothesis is valid, the properties of that circumstellar dust are very different from its interstellar counterpart.

We developed decade-long NIR template light curves for 658 LMC Miras using the $I$-band OGLE-\rom{3} light curves, and derived PLRs based on these templates. We found that the observed magnitudes of the C-rich Miras are fainter than those predicted from O-rich Miras, and these differences decrease with increasing wavelength. The magnitude differences are highly correlated across $JHK_s$ bands. These facts support our hypothesis of circumstellar extinction for the C-rich Miras.

However, our hypothesis is not fully justified within this work, as we cannot rule out the possibility that the circumstellar extinction of C-rich Miras is coupled with the intrinsic color variations from the stars and/or the dust envelopes. If the intrinsic color variations exist, the aforementioned slope of C-rich Mira color indices is no longer the color excess ratio, and its comparison with interstellar extinction laws becomes invalid.

As part of the hypothesis, we assumed that the circumstellar extinction for O-rich Miras is marginal in the NIR. The low scatter in the PLRs of O-rich Miras supports this assumption. If the circumstellar dust around O-rich Miras were significant, we would expect extinction variations would increase the dispersion in those relations. Future studies of O-rich Mira PLRs in other systems with known distances will be even more helpful to check the amount of circumstellar extinction for O-rich Miras.

\acknowledgements
W.Y. and L.M.M. acknowledge financial support from NSF grant AST-1211603 and from the Mitchell Institute for Fundamental Physics and Astronomy at Texas A\&M University.  S.H. was partially supported by the Texas A\&M University-NSFC Joint Research Program. JZH was partially supported by NSF grant DMS-1208952. CCN thanks the funding from Ministry of Science and Technology (Taiwan) under the contract 104-2112-M-008-012-MY3. SMK thanks SUNY-Oswego for startup funds that funded much of the initial telescope time for LMCNISS. We thank the referee for valuable suggestions that improved this manuscript.

\bibliographystyle{aasjournal}
\bibliography{./lmcnir}

\begin{thebibliography}{}
\expandafter\ifx\csname natexlab\endcsname\relax\def\natexlab#1{#1}\fi
\providecommand{\url}[1]{\href{#1}{#1}}

\bibitem[{{Alcock} {et~al.}(1993){Alcock}, {Allsman}, {Axelrod}, {Bennett},
  {Cook}, {Park}, {Marshall}, {Stubbs}, {Griest}, {Perlmutter}, {Sutherland},
  {Freeman}, {Peterson}, {Quinn}, \& {Rodgers}}]{1993ASPC...43..291A}
{Alcock}, C., {Allsman}, R.~A., {Axelrod}, T.~S., {et~al.} 1993, in
  Astronomical Society of the Pacific Conference Series, Vol.~43, Sky Surveys.
  Protostars to Protogalaxies, ed. B.~T. {Soifer}, 291

\bibitem[{{Aringer} {et~al.}(2009){Aringer}, {Girardi}, {Nowotny}, {Marigo}, \&
  {Lederer}}]{2009A&A...503..913A}
{Aringer}, B., {Girardi}, L., {Nowotny}, W., {Marigo}, P., \& {Lederer}, M.~T.
  2009, \aap, 503, 913

\bibitem[{{Bernal} {et~al.}(2016){Bernal}, {Verde}, \&
  {Riess}}]{2016JCAP...10..019B}
{Bernal}, J.~L., {Verde}, L., \& {Riess}, A.~G. 2016, \jcap, 10, 019

\bibitem[{{Bladh} {et~al.}(2015){Bladh}, {H{\"o}fner}, {Aringer}, \&
  {Eriksson}}]{2015A&A...575A.105B}
{Bladh}, S., {H{\"o}fner}, S., {Aringer}, B., \& {Eriksson}, K. 2015, \aap,
  575, A105

\bibitem[{{Carpenter}(2001)}]{2001AJ....121.2851C}
{Carpenter}, J.~M. 2001, \aj, 121, 2851

\bibitem[{{Eggen}(1975)}]{1975ApJ...195..661E}
{Eggen}, O.~J. 1975, \apj, 195, 661

\bibitem[{{Feast}(2009)}]{2009asrp.proc...48F}
{Feast}, M.~W. 2009, in AGB Stars and Related Phenomena, ed. T.~{Ueta},
  N.~{Matsunaga}, \& Y.~{Ita}, 48

\bibitem[{{Feast} {et~al.}(1989){Feast}, {Glass}, {Whitelock}, \&
  {Catchpole}}]{1989MNRAS.241..375F}
{Feast}, M.~W., {Glass}, I.~S., {Whitelock}, P.~A., \& {Catchpole}, R.~M. 1989,
  \mnras, 241, 375

\bibitem[{{Feast} {et~al.}(1982){Feast}, {Robertson}, {Catchpole}, {Evans},
  {Glass}, \& {Carter}}]{1982MNRAS.201..439F}
{Feast}, M.~W., {Robertson}, B.~S.~C., {Catchpole}, R.~M., {et~al.} 1982,
  \mnras, 201, 439

\bibitem[{{Fouqu{\'e}} {et~al.}(2007){Fouqu{\'e}}, {Arriagada}, {Storm},
  {Barnes}, {Nardetto}, {M{\'e}rand}, {Kervella}, {Gieren}, {Bersier},
  {Benedict}, \& {McArthur}}]{2007A&A...476...73F}
{Fouqu{\'e}}, P., {Arriagada}, P., {Storm}, J., {et~al.} 2007, \aap, 476, 73

\bibitem[{{Gerasimovic}(1928)}]{1928PNAS...14..963G}
{Gerasimovic}, B.~P. 1928, Proceedings of the National Academy of Science, 14,
  963

\bibitem[{{Glass} {et~al.}(1987){Glass}, {Catchpole}, {Feast}, {Whitelock}, \&
  {Reid}}]{1987ASSL..132...51G}
{Glass}, I.~S., {Catchpole}, R.~M., {Feast}, M.~W., {Whitelock}, P.~A., \&
  {Reid}, I.~N. 1987, in Astrophysics and Space Science Library, Vol. 132, Late
  Stages of Stellar Evolution, ed. S.~{Kwok} \& S.~R. {Pottasch}, 51--54

\bibitem[{{Glass} \& {Feast}(1982)}]{1982MNRAS.199..245G}
{Glass}, I.~S., \& {Feast}, M.~W. 1982, \mnras, 199, 245

\bibitem[{{Glass} \& {Lloyd Evans}(1981)}]{1981Natur.291..303G}
{Glass}, I.~S., \& {Lloyd Evans}, T. 1981, \nat, 291, 303

\bibitem[{{Glass} {et~al.}(1995){Glass}, {Whitelock}, {Catchpole}, \&
  {Feast}}]{1995MNRAS.273..383G}
{Glass}, I.~S., {Whitelock}, P.~A., {Catchpole}, R.~M., \& {Feast}, M.~W. 1995,
  \mnras, 273, 383

\bibitem[{{Gyllenberg}(1929)}]{1929MeLuS..53....3G}
{Gyllenberg}, W. 1929, Meddelanden fran Lunds Astronomiska Observatorium Serie
  II, 53, 3

\bibitem[{{He} {et~al.}(2016){He}, {Yuan}, {Huang}, {Long}, \&
  {Macri}}]{2016AJ....152..164H}
{He}, S., {Yuan}, W., {Huang}, J.~Z., {Long}, J., \& {Macri}, L.~M. 2016, \aj,
  152, 164

\bibitem[{{H{\"o}fner} {et~al.}(2016){H{\"o}fner}, {Bladh}, {Aringer}, \&
  {Ahuja}}]{2016A&A...594A.108H}
{H{\"o}fner}, S., {Bladh}, S., {Aringer}, B., \& {Ahuja}, R. 2016, \aap, 594,
  A108

\bibitem[{{Ita} \& {Matsunaga}(2011)}]{2011MNRAS.412.2345I}
{Ita}, Y., \& {Matsunaga}, N. 2011, \mnras, 412, 2345

\bibitem[{{Macri} {et~al.}(2015){Macri}, {Ngeow}, {Kanbur}, {Mahzooni}, \&
  {Smitka}}]{2015AJ....149..117M}
{Macri}, L.~M., {Ngeow}, C.-C., {Kanbur}, S.~M., {Mahzooni}, S., \& {Smitka},
  M.~T. 2015, \aj, 149, 117

\bibitem[{{Madore} \& {Freedman}(1991)}]{1991PASP..103..933M}
{Madore}, B.~F., \& {Freedman}, W.~L. 1991, \pasp, 103, 933

\bibitem[{{Marigo} {et~al.}(2013){Marigo}, {Bressan}, {Nanni}, {Girardi}, \&
  {Pumo}}]{2013MNRAS.434..488M}
{Marigo}, P., {Bressan}, A., {Nanni}, A., {Girardi}, L., \& {Pumo}, M.~L. 2013,
  \mnras, 434, 488

\bibitem[{{Marigo} {et~al.}(2017){Marigo}, {Girardi}, {Bressan}, {Rosenfield},
  {Aringer}, {Chen}, {Dussin}, {Nanni}, {Pastorelli}, {Rodrigues}, {Trabucchi},
  {Bladh}, {Dalcanton}, {Groenewegen}, {Montalb{\'a}n}, \&
  {Wood}}]{2017ApJ...835...77M}
{Marigo}, P., {Girardi}, L., {Bressan}, A., {et~al.} 2017, \apj, 835, 77

\bibitem[{{Matsuura} {et~al.}(2009){Matsuura}, {Barlow}, {Zijlstra},
  {Whitelock}, {Cioni}, {Groenewegen}, {Volk}, {Kemper}, {Kodama}, {Lagadec},
  {Meixner}, {Sloan}, \& {Srinivasan}}]{2009MNRAS.396..918M}
{Matsuura}, M., {Barlow}, M.~J., {Zijlstra}, A.~A., {et~al.} 2009, \mnras, 396,
  918

\bibitem[{{Nanni} {et~al.}(2013){Nanni}, {Bressan}, {Marigo}, \&
  {Girardi}}]{2013MNRAS.434.2390N}
{Nanni}, A., {Bressan}, A., {Marigo}, P., \& {Girardi}, L. 2013, \mnras, 434,
  2390

\bibitem[{{Nishiyama} {et~al.}(2009){Nishiyama}, {Tamura}, {Hatano}, {Kato},
  {Tanab{\'e}}, {Sugitani}, \& {Nagata}}]{2009ApJ...696.1407N}
{Nishiyama}, S., {Tamura}, M., {Hatano}, H., {et~al.} 2009, \apj, 696, 1407

\bibitem[{{Pietrzy{\'n}ski} {et~al.}(2013){Pietrzy{\'n}ski}, {Graczyk},
  {Gieren}, {Thompson}, {Pilecki}, {Udalski}, {Soszy{\'n}ski}, {Koz{\l}owski},
  {Konorski}, {Suchomska}, {Bono}, {Moroni}, {Villanova}, {Nardetto},
  {Bresolin}, {Kudritzki}, {Storm}, {Gallenne}, {Smolec}, {Minniti}, {Kubiak},
  {Szyma{\'n}ski}, {Poleski}, {Wyrzykowski}, {Ulaczyk}, {Pietrukowicz},
  {G{\'o}rski}, \& {Karczmarek}}]{2013Natur.495...76P}
{Pietrzy{\'n}ski}, G., {Graczyk}, D., {Gieren}, W., {et~al.} 2013, \nat, 495,
  76

\bibitem[{{Planck Collaboration} {et~al.}(2016){Planck Collaboration}, {Ade},
  {Aghanim}, {Arnaud}, {Ashdown}, {Aumont}, {Baccigalupi}, {Banday},
  {Barreiro}, {Bartlett}, \& et~al.}]{2016A&A...594A..13P}
{Planck Collaboration}, {Ade}, P.~A.~R., {Aghanim}, N., {et~al.} 2016, \aap,
  594, A13

\bibitem[{{Riess} {et~al.}(2016){Riess}, {Macri}, {Hoffmann}, {Scolnic},
  {Casertano}, {Filippenko}, {Tucker}, {Reid}, {Jones}, {Silverman},
  {Chornock}, {Challis}, {Yuan}, {Brown}, \& {Foley}}]{2016ApJ...826...56R}
{Riess}, A.~G., {Macri}, L.~M., {Hoffmann}, S.~L., {et~al.} 2016, \apj, 826, 56

\bibitem[{{Robertson} \& {Feast}(1981)}]{1981MNRAS.196..111R}
{Robertson}, B.~S.~C., \& {Feast}, M.~W. 1981, \mnras, 196, 111

\bibitem[{{Soszynski} {et~al.}(2004{\natexlab{a}}){Soszynski}, {Udalski},
  {Kubiak}, {Szymanski}, {Pietrzynski}, {Zebrun}, {Szewczyk}, \&
  {Wyrzykowski}}]{2004AcA....54..129S}
{Soszynski}, I., {Udalski}, A., {Kubiak}, M., {et~al.} 2004{\natexlab{a}},
  \actaa, 54, 129

\bibitem[{{Soszynski} {et~al.}(2004{\natexlab{b}}){Soszynski}, {Udalski},
  {Kubiak}, {Szymanski}, {Pietrzynski}, {Zebrun}, {Szewczyk}, {Wyrzykowski}, \&
  {Dziembowski}}]{2004AcA....54..347S}
---. 2004{\natexlab{b}}, \actaa, 54, 347

\bibitem[{{Soszynski} {et~al.}(2007){Soszynski}, {Dziembowski}, {Udalski},
  {Kubiak}, {Szymanski}, {Pietrzynski}, {Wyrzykowski}, {Szewczyk}, \&
  {Ulaczyk}}]{2007AcA....57..201S}
{Soszynski}, I., {Dziembowski}, W.~A., {Udalski}, A., {et~al.} 2007, \actaa,
  57, 201

\bibitem[{{Soszy{\'n}ski} {et~al.}(2009){Soszy{\'n}ski}, {Udalski},
  {Szyma{\'n}ski}, {Kubiak}, {Pietrzy{\'n}ski}, {Wyrzykowski}, {Szewczyk},
  {Ulaczyk}, \& {Poleski}}]{2009AcA....59..239S}
{Soszy{\'n}ski}, I., {Udalski}, A., {Szyma{\'n}ski}, M.~K., {et~al.} 2009,
  \actaa, 59, 239

\bibitem[{Tibshirani(1994)}]{Tibshirani1994}
Tibshirani, R. 1994, Journal of the Royal Statistical Society, Series B, 58,
  267

\bibitem[{{Udalski} {et~al.}(2008){Udalski}, {Szymanski}, {Soszy{\'n}ski}, \&
  {Poleski}}]{2008AcA....58...69U}
{Udalski}, A., {Szymanski}, M.~K., {Soszy{\'n}ski}, I., \& {Poleski}, R. 2008,
  \actaa, 58, 69

\bibitem[{{van Loon} {et~al.}(1998){van Loon}, {Zijlstra}, {Whitelock}, {te
  Lintel Hekkert}, {Chapman}, {Loup}, {Groenewegen}, {Waters}, \&
  {Trams}}]{1998A&A...329..169V}
{van Loon}, J.~T., {Zijlstra}, A.~A., {Whitelock}, P.~A., {et~al.} 1998, \aap,
  329, 169

\bibitem[{{Wang} \& {Jiang}(2014)}]{2014ApJ...788L..12W}
{Wang}, S., \& {Jiang}, B.~W. 2014, \apjl, 788, L12

\bibitem[{{Whitelock} \& {Feast}(2014)}]{2014EAS....67..263W}
{Whitelock}, P.~A., \& {Feast}, M.~W. 2014, in EAS Publications Series,
  Vol.~67, EAS Publications Series, 263--269

\bibitem[{{Whitelock} {et~al.}(2006){Whitelock}, {Feast}, {Marang}, \&
  {Groenewegen}}]{2006MNRAS.369..751W}
{Whitelock}, P.~A., {Feast}, M.~W., {Marang}, F., \& {Groenewegen}, M.~A.~T.
  2006, \mnras, 369, 751

\bibitem[{{Whitelock} {et~al.}(2008){Whitelock}, {Feast}, \& {van
  Leeuwen}}]{2008MNRAS.386..313W}
{Whitelock}, P.~A., {Feast}, M.~W., \& {van Leeuwen}, F. 2008, \mnras, 386, 313

\bibitem[{{Whitelock} {et~al.}(2003){Whitelock}, {Feast}, {van Loon}, \&
  {Zijlstra}}]{2003MNRAS.342...86W}
{Whitelock}, P.~A., {Feast}, M.~W., {van Loon}, J.~T., \& {Zijlstra}, A.~A.
  2003, \mnras, 342, 86

\bibitem[{{Wilson} \& {Merrill}(1942)}]{1942ApJ....95..248W}
{Wilson}, R.~E., \& {Merrill}, P.~W. 1942, \apj, 95, 248

\bibitem[{{Wood}(1990)}]{1990ASPC...11..355W}
{Wood}, P.~R. 1990, in Astronomical Society of the Pacific Conference Series,
  Vol.~11, Confrontation Between Stellar Pulsation and Evolution, ed.
  C.~{Cacciari} \& G.~{Clementini}, 355--363

\bibitem[{{Wood}(2000)}]{2000PASA...17...18W}
{Wood}, P.~R. 2000, \pasa, 17, 18

\bibitem[{{Wood}(2015)}]{2015MNRAS.448.3829W}
---. 2015, \mnras, 448, 3829

\bibitem[{{Wood} {et~al.}(1983){Wood}, {Bessell}, \&
  {Fox}}]{1983ApJ...272...99W}
{Wood}, P.~R., {Bessell}, M.~S., \& {Fox}, M.~W. 1983, \apj, 272, 99

\bibitem[{{Wood} {et~al.}(1999){Wood}, {Alcock}, {Allsman}, {Alves}, {Axelrod},
  {Becker}, {Bennett}, {Cook}, {Drake}, {Freeman}, {Griest}, {King}, {Lehner},
  {Marshall}, {Minniti}, {Peterson}, {Pratt}, {Quinn}, {Stubbs}, {Sutherland},
  {Tomaney}, {Vandehei}, \& {Welch}}]{1999IAUS..191..151W}
{Wood}, P.~R., {Alcock}, C., {Allsman}, R.~A., {et~al.} 1999, in IAU Symposium,
  Vol. 191, Asymptotic Giant Branch Stars, ed. T.~{Le Bertre}, A.~{Lebre}, \&
  C.~{Waelkens}, 151

\bibitem[{{Yuan} {et~al.}(2017){Yuan}, {He}, {Macri}, {Long}, \&
  {Huang}}]{2017AJ....153..170Y}
{Yuan}, W., {He}, S., {Macri}, L.~M., {Long}, J., \& {Huang}, J.~Z. 2017, \aj,
  153, 170

\end{thebibliography}
\end{document}